\documentclass[12pt]{article}

\input epsf
\usepackage{epsfig,graphicx,epsf,hyperref}
\usepackage{amsmath,amsfonts,amssymb,amsbsy}
\hypersetup{colorlinks=true,linkcolor=blue,citecolor=blue,
urlcolor=blue}

\def\i{\mbox{i}}
\def\\i{\mbox{\scriptsize{i}}}
\def\d{\mbox{d}}
\def\Im{\mbox{Im}}

\title{Coulomb-nuclear interference in elastic proton scattering in the eikonal approach}
\author{M. L. Nekrasov\\
{\small\it 
Institute for High Energy Physics, NRC ``Kurchatov
Institute'',}  \vspace*{-4\baselineskip}\\
{\small\it Protvino 142281, Russia} }

\date{}

\begin{document}
\maketitle

\begin{abstract}
We find exact solution in the Cahn eikonal model, which describes Coulomb-nuclear interference in elastic scattering of charged hadrons. The cases of both point-like and extended particles equipped with electromagnetic form factors are considered. According to the solution obtained the Coulomb-nuclear contributions are not exponentiated and cannot be added to the Coulomb phase. At the same time, the $O(\alpha)$-approximation of the amplitude is ambiguous, which makes it unsuitable for data processing.
\end{abstract} 

\section{Introduction}\label{sec1}

The Coulomb-nuclear interference (CNI) is a unique source of information about the hadronic part of the amplitude of proton elastic scattering near the forward direction. In this regard, the most accurate calculation of electromagnetic contributions to the amplitude at low momentum transfers is an important (theoretical) component of the study of elastic proton scattering at modern colliders.

In fact, there is an extensive literature on this issue, including the original studies \cite{Bethe,Soloviev,WY,Cahn} and subsequent developments \cite{KL,Petrov2018,Durand2020} (see also references therein). Apparently, the first attempt to describe the combined Coulomb and nuclear contributions to the elastic scattering amplitude was made in \cite{Bethe} with the help of WKB method used in the potential theory. Subsequently, when studying the problem in the quantum field theory (QFT), the analysis in any case relied on theorems on the elimination of IR divergences in the cross section. Recall that according to them the IR-divergent contributions to the amplitude are factorized. Some of them have the form of an exponential with a complex phase, and they do not make contributions to the cross section. The rest are canceled by IR-divergent contributions from undetected real soft photons. After eliminating the IR divergencies, the finite soft photon contributions give different contributions to the purely Coulomb and Coulomb-nuclear parts of the amplitude, leading to a non-trivial CNI effect \cite{Soloviev}. 

A detailed analysis of CNI based on Feynman diagrams in the leading non-trivial order of power expansion in the fine structure constant $\alpha$ was carried out in \cite{WY}. It was shown that after eliminating IR-divergent contributions, the ``imaginary'', if attributed to the exponential argument, contributions of soft photons do not coincide in the pure Coulomb and Coulomb nuclear contributions. Moreover, they arise only in the diagrams with virtual photons connecting the legs of different external particles. At the same time, the ``real'' (in the above sense) parts of the soft photon contributions are common for all parts of the amplitude and insignificant in the CNI region. 

Further, the analysis of CNI was carried out in the eikonal approach, namely in the model with one-photon exchange in the Born approximation  \cite{Cahn}. In fact, this model determines the high-energy behavior of the amplitude with limited momentum transfer precisely by collecting contributions of photon exchanges between the legs of different external particles \cite{Cheng-Wu,Levy-Sucher,Collins}. Accordingly, the result of \cite{WY} was reproduced in \cite{Cahn}. An advantage of the eikonal approach is also that it allows one to consistently take into account the electromagnetic form factors in the case of  scattering of spatially extended particles. A corresponding generalization of the amplitude was found in \cite{Cahn}. Subsequently, it was transformed into a form better suited for practical applications \cite{KL} and then used in processing TOTEM data \cite{TOTEM8}, see also next TOTEM~publications.

However, the potential opportunity of the eikonal approach was not fully realized in \cite{Cahn,KL}, since the final result was obtained only in the leading non-trivial order in $\alpha$. Moreover, the derivation of the final formulas was not entirely correct. In this paper we eliminate this drawback and find a generalization that takes into account the contributions to all orders in $\alpha$. Actually, we find an exact solution in this version of the eikonal model. The main means that made it possible to obtain this result is the consistent application of the analytical regularization method.

The structure of this paper is as follows. In the next section we find exact solution in the eikonal model \cite{Cahn} in the case of scattering of point-like particles. Section \ref{sec3} generalizes the results to the case of spatially extended particles. In the final section, we discuss the results.

\section{Scattering of point-like particles}\label{sec2}

Following the notation of \cite{Cahn}, we consider the amplitude of elastic scattering of charged particles (protons) in the normalization
\begin{equation}\label{R1}
\frac{\d \sigma}{\d t} = \frac{\pi}{s p^2_{cm}} |F|^2.
\end{equation} 
At high energies and small scattering angles the amplitude may be represented in the eikonal parameterization,
\begin{equation}\label{R2}
F(s,q^2) = 
\frac{s}{4\pi \i} \int \d^2 {\bf b} \;\; 
e^{\\i {\bf q b}} \left[ e^{2\\i\delta(s,{\bf b})} - 1 \right].
\end{equation}
Here $s$ is the square of the center-of-mass energy, ${\bf q}$ is the momentum transfer, $t=-{\bf q}^2$, $q = |{\bf q}|$, ${\bf b}$ is the impact parameter, and $\delta(s,{\bf b})$ is the phase shift. 

In the eikonal model, the phase shift $\delta(s,b)$ is determined by the Born amplitude,
\begin{equation}\label{R3}
\delta(s,b) = 
\frac{2\pi}{s} \int \frac{\d^2 {\bf q}}{(2\pi)^2} \; 
e^{-\\i {\bf q b}} F_{\mbox{\scriptsize Born}}(s,{\bf q}^2) \,.
\end{equation}
So the choice of $F_{\mbox{\scriptsize Born}}(s,{\bf q}^2)$ determines the particular model under consideration. In the potential model, the Born amplitude is determined by the scattering potential. In the Regge theory and QFT, $F_{\mbox{\scriptsize Born}}$ is defined as the simplest amplitude with exchange in the $t$-channel. Formula (\ref{R2}) in this case describes the result of summing the generalized ladder of similar exchanges in the approximation of large $s$ and limited $t$ \cite{Cheng-Wu,Levy-Sucher,Collins}. Below we follow the QFT version of the eikonal model, in which Coulomb interaction occurs due to photon exchange.

\subsection{Coulomb scattering}\label{sec2.1}

In the case of Coulomb scattering of point-like particles with identical electric charges (generalization to the case of different charges is trivial) we define, following \cite{Cahn}, the Born amplitude as follows\footnote{Thus, the vacuum polarization contributions and radiative corrections to the vertex functions are ignored. According to \cite{WY}, this leads to a loss in the accuracy by about half a percent of the $O(\alpha)$ contribution in the ${\bf q}^2$ range of interest.} 
\begin{equation}\label{R4}
F^C_{\mbox{\scriptsize Born}} = 
- \frac{\alpha s}{{\bf q}^2 + \lambda^2}\,.
\end{equation} 
Here $\lambda$ is the fictitious photon mass introduced to regularize the IR divergences. Substituting (\ref{R4}) into (\ref{R3}), we get
\begin{equation}\label{R5}
\delta^{C}(b) = 
\frac{1}{2\pi} \int \d^2 {\bf q} \; 
e^{-\\i {\bf q b}} \frac{-\alpha}{{\bf q}^2 + \lambda^2} =
- \alpha \mbox{K}_0 (b \lambda)\,.
\end{equation}
Here $\mbox{K}_0 (z)$ is the MacDonald function of the zero order. So the scattering amplitude of identical elementary charges is
\begin{equation}\label{R6}
F^{C}(s,q^2) = 
\frac{s}{4\pi \i} \int \d^2 {\bf b} \;\; 
e^{\\i {\bf q b}} \left[ e^{2\\i\delta^{C}(b)} - 1 \right] .
\end{equation}
Note at once that for $z \to \infty$ the MacDonald function decreases exponentially, $\mbox{K}_0 (z) \simeq \exp(-z) \sqrt{ \pi/(2z) }$, and for small $z$ it has the behavior ($\gamma = 0.5772\dots$)
\begin{equation}\label{R7}
\mbox{K}_0 (z) = -\ln(z/2) - \gamma + \bar{o}(z)\,.
\end{equation} 
Hence, for finite $\lambda$ and $\Im\,\alpha < 1$ integral (\ref{R6}) converges and defines an analytic function of $\alpha$. In particular, with $\alpha \to 0$ we have $F^{C} = F^C_{\mbox{\scriptsize Born}} + O(\alpha^2)$. 

Our task is to isolate into a factor the singularity at $\lambda\to 0$ and set $\lambda=0$ in the remaining part of the amplitude. Unfortunately, if we substitute into (\ref{R6}) the asymptotics of $\mbox{K}_0 (b \lambda)$ at $\lambda \to 0$, we obtain an integral that diverges for large $b$. In this regard, consider the amplitude at $1/4 < \Im\,\alpha < 1$. Then formula (\ref{R6}) at $\lambda\to 0$ can be transformed by virtue of (\ref{R7}) to 
\begin{equation}\label{R8}
F^{C}(s,q^2){\bigl|_{\lambda \to 0 }\bigr.} = 
\frac{s}{2 \i} \! \int_0^{\infty} \!\! b \d b \; J_0 (qb) \; (bq)^{2\\i \alpha}
\left(\frac{\lambda e^{\gamma}}{2q} \right)^{\!\!2\\i \alpha} \!\!  + \i s \pi \delta({\bf q}) .
\end{equation}
Here $J_0 (z)$ is the Bessel function of zero order. Direct calculation of the integral gives
\begin{equation}\label{R9}
F^{C}(s,q^2){\bigl|_{\lambda \to 0 }\bigr.} = 
- \, \frac{s \alpha}{q^2} \left(\frac{\lambda e^{\gamma}}{q}  \right)^{\!\!2\\i \alpha} \frac{\Gamma(1+\i \alpha)}{\Gamma(1-\i \alpha)} + \i s \pi \delta({\bf q})\,.
\end{equation}
This can be also written as
\begin{equation}\label{R10}
F^{C}(s,q^2){\bigl|_{\lambda \to 0 }\bigr.} = 
-\,\frac{s \alpha}{q^2} \, e^{\\i \alpha\ln (\lambda^2/q^2) \,+\, \\i \Phi(\alpha)} + \i s \pi \delta({\bf q})\,,
\end{equation}
where
\begin{equation}\label{R11}
\i \Phi(\alpha) = \i 2 \alpha\gamma +
\ln \frac{\Gamma(1+\i\alpha)}{\Gamma(1-\i \alpha)}\,.
\end{equation}

In formula (\ref{R10}) we can do the inverse analytic continuation and return to the real $\alpha$. In this case $[\i \alpha\ln (\lambda^2/q^2) + \i \Phi]$ becomes purely complex, i.e.~determines a phase that accumulates IR-divergent contributions in accordance with general theorems.\footnote{In what follows, we omit constant term $\Phi(\alpha)$, assuming that it is absorbed by redefined $\lambda$.} The occurrence of the phase is a consequence of the summation of the generalized ladder of Born contributions. At ${\bf q} \not= 0$ the phase can be discarded, and (\ref{R10}) gives the desired result. 

However the presence of the $\delta$-function in (\ref{R10}) is not incidental, and it is necessary for the correct derivation of subsequent formulas (this $\delta$-function was overlooked in \cite{Cahn}). Its origin is due to the term $-1$ in square brackets in (\ref{R6}), and its role is to cancel out the unscattered wave. To see the presence of this wave in the first term in (\ref{R10}), one should send $\alpha \to 0$ in this term. In fact, this operation is non-trivial due to the singular behavior of the phase at $q^2 = 0$. To do this correctly, we use the formula
\begin{equation}\label{R12}
|\tilde{\bf q}|^{-2+\epsilon} = \frac{2\pi}{\epsilon}\delta(\tilde{\bf q}) + |\tilde{\bf q}|_{+}^{-2} + O(\epsilon)
\end{equation}  
that takes place in the analytic regularization approach \cite{Gelfand}. Here $\tilde{\bf q}$ is a dimensionless vector in 2-dimensional space, and the second term is an associated generalized function (distribution) of first order and of degree $-2$ (at $\tilde{\bf q} \not= 0$ coincides with the ordinary function $|\tilde{\bf q}|^{-2}$). In our case $\tilde{\bf q} = {\bf q}/\lambda$, and by virtue of (\ref{R12}) the first term in (\ref{R10})~at $\alpha\to 0$ turns into $-\i s\pi \delta({\bf q})$, which is the non-scattered wave contribution. So, the expansion of (\ref{R10}) in powers of $\alpha$ for all ${\bf q}$ looks like $F^{C}(s,q^2){\bigl|_{\lambda \to 0 }\bigr.} =\hat{F}^C_{\mbox{\scriptsize Born}} + O(\alpha^2)$, where the first term is of order $O(\alpha)$ and in the neighborhood of ${\bf q} = 0$ is a distribution. This is not exactly the same as the expansion of (\ref{R6}) in powers of $\alpha$, since out of the phase formula (\ref{R10}) contains only contributions at $\lambda =0$.

\subsection{Coulomb-nuclear scattering}\label{sec2.2}

If scattering occurs due to both the Coulomb and strong interactions, the amplitude in the eikonal model \cite{Cahn} is written as
\begin{equation}\label{R13}
F^{N+C}(s,q^2) = 
\frac{s}{4\pi \i} \int \d^2 {\bf b} \;\; 
e^{\\i {\bf q b}} \left[ e^{2\\i \left( \delta^{C}+\delta^{N} \right)} - 1 \right] .
\end{equation}
Here $\delta^{N}=\delta^{N}(s,b)$ is a phase shift due to the strong interactions. Accordingly, with the Coulomb interaction switched off, the amplitude of purely strong interactions is
\begin{equation}\label{R14}
F^{N}(s,q^2) = 
\frac{s}{4\pi \i} \int \d^2 {\bf b} \;\; 
e^{\\i {\bf q b}} \left[ e^{2\\i \delta^{N}(b)} - 1 \right],
\end{equation}
which is assumed to be an analytic function of $q^2$. Combining (\ref{R6}), (\ref{R13}), and (\ref{R14}), one can obtain \cite{Cahn} 
\begin{equation}\label{R15}
F^{N+C}(s,q^2) = F^{C}(s,q^2) + F^{N}(s,q^2) 
+ \frac{i}{\pi s} \! \int \!\d^2 {\bf q}' \; 
F^{C}(s,q'^{\,2}) \, F^{N}(s,[{\bf q-q}']^2) \,. 
\end{equation}
Considering this formula at $1/4 < \Im\,\alpha < 1$ and substituting (\ref{R10}), we get at $\lambda\to 0$ and ${\bf q} \not= 0$
\begin{equation}\label{R16}
F^{N+C}(s,q^2) \; e^{-{\mbox{\scriptsize{i}}} \alpha\ln (\lambda^2/q^2)} = - \, \frac{s \alpha}{q^2} 
+\frac{i}{\pi s} \! \int \!\d^2 {\bf q}' \;
\left[\frac{- s \alpha}{q'^{\,2}}\right]
\left(\frac{q^{\,2}}{q'^{\,2}}\right)^{{\mbox{\scriptsize{i}}} \alpha} 
F^{N}(s,[{\bf q-q}']^2) \,. 
\end{equation}
When passing to (\ref{R16}), we took into account that $F^{C}(s,q'^{\,2}){\bigl|_{\lambda \to 0 }\bigr.}$ contains $\delta$-function, which cancels out $F^{N}(s,q^2)$ in (\ref{R15}). At the same time, the $\delta$-function in $F^{C}(s,q^2){\bigl|_{\lambda \to 0 }\bigr.}$ outside the integral does not make contribution at ${\bf q}\not= 0$. We also move the phase of both $F^{C}$ to the l.h.s., simultaneously introducing the correction in the integrand. Next, for convenience of working with the singularity at $q'^{\,2} = 0$, we rewrite the r.h.s.~(\ref{R16}) in the form 
\begin{eqnarray}\label{R17}
& \displaystyle
\mbox{r.h.s.~(\ref{R16})} = \;-\; \frac{s \alpha}{q^2} 
\;+\; \frac{i}{\pi s} \! \int^{s} \!\d^2 {\bf q}' \;
\left[\frac{- s \alpha}{q'^{\,2}}\right]
\left(\frac{q^{\,2}}{q'^{\,2}}\right)^{\!\mbox{\scriptsize{i}} \alpha} F^{N}(s,q^2) & \nonumber\\[0.5\baselineskip]
& \displaystyle + \;\frac{i}{\pi s} \! \int^{s} \!\d^2 {\bf q}' \;
\left[\frac{- s \alpha}{q'^{\,2}}\right]
\left(\frac{q^{\,2}}{q'^{\,2}}\right)^{\!\mbox{\scriptsize{i}} \alpha} 
\left[F^{N}(s,[{\bf q-q}']^2) - F^{N}(s,q^2) \right]. &
\end{eqnarray} 
Following \cite{Cahn}, we indicate here the upper limit of integration as a reminder that the square of the transfer is limited, $q^2 = 2 p^2 (1 - \cos\theta) < s$. For the above values of $\Im\,\alpha$, the first integral in (\ref{R17}) converges, and after its direct calculation we get
\begin{eqnarray}\label{R18}
& \displaystyle
F^{N+C}(s,q^2) \; e^{-\mbox{\scriptsize{i}} \alpha\ln (\lambda^2/q^2)} = - \, \frac{s \alpha}{q^2} + \left(\frac{s}{q^2}\right)^{\!- \mbox{\scriptsize{i}} \alpha} \! F^{N}(s,q^2)
& \nonumber\\[0.5\baselineskip]
& \displaystyle 
- \; \frac{i \alpha}{\pi} \! \int^{s} \! \;
\frac{\d^2 {\bf q}'}{q'^{\,2}}
\left(\frac{q^{\,2}}{q'^{\,2}}\right)^{\!\mbox{\scriptsize{i}} \alpha} 
\left[F^{N}(s,[{\bf q-q}']^2) - F^{N}(s,q^2) \right] .&
\end{eqnarray}
In (\ref{R18}), we can return to the real $\alpha$, and this formula represents an exact solution for the scattering amplitude in the eikonal model \cite{Cahn}. 

Note, however, that after eliminating the IR divergences, the common overall complex phase is not uniquely determined. From the point of view of physical interpretation, a more preferable option is obtained by multiplying (\ref{R18}) by a factor $\left( s/q^2 \right)^{\mbox{\scriptsize{i}} \alpha}$:
\begin{eqnarray}\label{R19}
& \displaystyle
F^{N+C}(s,q^2) \; e^{-\mbox{\scriptsize{i}} \alpha\ln (\lambda^2/s)} = - \, \frac{s \alpha}{q^2}\; 
e^{\mbox{\scriptsize{i}} \alpha\ln (s/q^2)} + F^{N}(s,q^2)
& \nonumber\\[0.5\baselineskip]
& \displaystyle 
- \; \frac{i \alpha}{\pi} \! \int^{s} \! \;
\frac{\d^2 {\bf q}'}{q'^{\,2}}
\left(\frac{s}{q'^{\,2}}\right)^{\!\mbox{\scriptsize{i}} \alpha} 
\left[F^{N}(s,[{\bf q-q}']^2) - F^{N}(s,q^2) \right] .&
\end{eqnarray}
In this formula, the first term is a purely Coulomb contribution with a singular phase at $q^2\to 0$. The second term describes purely hadronic contribution, and the third term is the Coulomb-nuclear contribution. It is noteworthy that in structure (\ref{R19}) coincides, and the phase in the Coulomb term exactly coincides with those previously obtained by Solov'ev \cite{Soloviev}.

Having the exact solution, we can obtain its expansion in powers of $\alpha$. However, since $F^{N}$ is independent of $\alpha$, the result depends on the choice of the overall complex phase in the exact solution. In particular, the expansion of (\ref{R18}) gives the result obtained earlier in \cite{Cahn},
\begin{eqnarray}\label{R20}
& \displaystyle
F^{N+C}(s,q^2) \; e^{-\mbox{\scriptsize{i}} \alpha\ln (\lambda^2/q^2)} = 
- \, \frac{s \alpha}{q^2} +  F^{N}(s,q^2) 
\left\{1 - \i \alpha \ln\frac{s}{q^2} \right.
& \nonumber\\[0.5\baselineskip]
& \displaystyle \left.
-\; \i \alpha \! \int_0^{s} \! \d q'^{\,2}\;
\frac{1}{|q^2-q'^{\,2}|} 
\left[\frac{F^{N}(s,q'^{\,2})}{F^{N}(s,q^2)} - 1 \right]
\right\} + O(\alpha^2)\,.
\end{eqnarray}
The expression in curly brackets was considered in \cite{Cahn} as the result of expansion of the phase factor, and with this interpretation it coincides with the result of West and Yennie \cite{WY}. However, in fact, the exact solution (\ref{R18}) suggests that only the second contribution in curly brackets in (\ref{R20}) is exponentiated. If we do expansion based on (\ref{R19}), we obtain 
\begin{eqnarray}\label{R21}
& \displaystyle
F^{N+C}(s,q^2) \; e^{-\mbox{\scriptsize{i}} \alpha\ln (\lambda^2/s)} =  
- \, \frac{s \alpha}{q^2}  
& \nonumber\\[0.2\baselineskip]
& \displaystyle 
+ \; F^{N}(s,q^2) 
\left\{1 -\; \i \alpha \! \int_0^{s} \! \d q'^{\,2}\;
\frac{1}{|q^2-q'^{\,2}|} 
\left[\frac{F^{N}(s,q'^{\,2})}{F^{N}(s,q^2)} - 1 \right] \right\}  + O(\alpha^2) . &
\end{eqnarray} 
This result differs from (\ref{R20}) by a term of the order of $O(\alpha)$. In the differential cross section this difference translates into a difference of the order of $O(\alpha^2)$. So to obtain the differential cross section with an accuracy of $O(\alpha^2)$, one should use the $O(\alpha^2)$-approximation of the amplitude. However, in turn, the latter approximation is determined ambiguously, and in the differential cross section this leads to the ambiguity in $O(\alpha^3)$ order, etc. Recall that in the exact solution this problem does not arise, since the complex phase in it is completely factorized.

\section{Account of the form factor}\label{sec3}

In the case of scattering of extended particles characterized by a spatial distribution, the initial formula (\ref{R6}) for the Coulomb contributions must be modified. Namely, still assuming that the Born approximation is determined by one-photon exchange, the summed generalized ladder of such contributions is the averaging of the expression
\begin{equation}\label{R22}
F^{C}(s,q^2;{\bf s}_a,{\bf s}_b) = 
\frac{s}{4\pi \i} \int \d^2 {\bf b} \;\; 
e^{\\i {\bf q b}} 
\left[ e^{2\\i\delta^{C}({\bf b-s}_a+{\bf s}_b)} - 1 \right].
\end{equation}
Here ${\bf s}_a$ and ${\bf s}_b$ define the positions of the points of  emission and absorption of the photon relative to the centers of scattering particles. The averaging means sandwiching (\ref{R22}) between the wave functions of initial and final states of scattering particles. Practically this means taking integrals $\d^2 {\bf s}_a$ and $\d^2{\bf s}_b$ with weight functions determining the charge distribution in scattering particles. Simultaneously, they define the probabilities that the emission and absorption of the photon occur at the points ${\bf s}_a$ and ${\bf s}_b$. So, the amplitude is determined as 
\begin{equation}\label{R23}
F^{C}(s,q^2) = 
\int  \d^2 {\bf s}_a \d^2 {\bf s}_b \;\,
\rho_a ({\bf s}_a) \rho_b ({\bf s}_b) \,
F^{C}(s,q^2;{\bf s}_a,{\bf s}_b),
\end{equation}
where the wight functions $\rho_\kappa ({\bf s}_\kappa)$, $\kappa = a,b$, are normalized to unity, $\int \d^2 {\bf s}_\kappa \, \rho_\kappa ({\bf s}_\kappa) =1$.

Next, we change the variables, $\{{\bf s}_a,{\bf s}_b\} \to \{{\bf s' - s''},{\bf -s''}\}$. Taking into account that $\rho_\kappa ({\bf s}_\kappa)$ are even functions, we obtain
\begin{equation}\label{R24}
F^{C}(s,q^2) = 
\frac{s}{4\pi \i} \int \!\d^2 {\bf b} \, 
e^{\\i {\bf q b}}   \int  \!\d^2 {\bf s'} \, 
e^{2\\i\delta^{C}({\bf b-s'})}
\int  \!\d^2 {\bf s''} \,\rho_a ({\bf s'-s''}) \rho_b ({\bf s''}) + \i s \pi \delta({\bf q})\,.
\end{equation}
The third integral in (\ref{R24}) with external ${\bf s'}$, is actually the Fourier of ${\cal F}_a(q^2){\cal F}_b(q^2)$, where ${\cal F}_\kappa(q^2)$ is the Fourier of $\rho_\kappa ({\bf s})$, i.e.~the electromagnetic form factor. Consequently, the second and third integrals taken together mean the double convolution of three functions. So their Fourier, defined by the first integral, gives the product of these functions in the momentum space. Taking this into account and taking into account (\ref{R6}) and (\ref{R10}), we get
\begin{equation}\label{R25}
F^{C}(s,q^2){\bigl|_{\lambda \to 0 }\bigr.} = - \, \frac{s \alpha}{q^2} \, e^{\\i \alpha\ln (\lambda^2/q^2)} \, {\cal F}_a(q^2){\cal F}_b(q^2) + \i s \pi \delta({\bf q})\,.
\end{equation}

When including strong interactions, we proceed similarly to section \ref{sec2.2} simultaneously assuming that the effects of spatial distribution are already taken into account in the nuclear amplitude $F^N$. Also assuming for simplicity ${\cal F}_a(q^2) = {\cal F}_b(q^2) \equiv {\cal F}(q^2)$, we obtain at $1/4 < \Im\,\alpha < 1$ the following generalization of formula (\ref{R17}),
\begin{eqnarray}\label{R26}
& \displaystyle
F^{N+C}(s,q^2) \; e^{-\mbox{\scriptsize{i}} \alpha\ln (\lambda^2/q^2)} = \;-\; \frac{s \alpha}{q^2} \; {\cal F}^2(q^2)
& \nonumber\\
& \displaystyle +\; \frac{i}{\pi s} \! \int \!\d^2 {\bf q}' \;
\left[\frac{- s \alpha}{q'^{\,2}}\right]
\left(\frac{q^{\,2}}{q'^{\,2}}\right)^{\!\mbox{\scriptsize{i}} \alpha}  {\cal F}^2(q'^{\,2}) \; F^{N}(s,q^2) 
& \nonumber\\[0.5\baselineskip]
& \displaystyle + \; \frac{i}{\pi s} \! \int \!\d^2 {\bf q}' \;
\left[\frac{- s \alpha}{q'^{\,2}}\right]
\left(\frac{q^{\,2}}{q'^{\,2}}\right)^{\!\mbox{\scriptsize{i}} \alpha}   {\cal F}^2(q'^{\,2})
\left[F^{N}(s,[{\bf q-q}']^2) - F^{N}(s,q^2) \right]. &
\end{eqnarray}  
In this case, we do not specify the limits of integration, since the presence of a rapidly decreasing form factor effectively cuts off the integrals before reaching  the upper limit in $q'^{\,2}$. Taking this into account, formula (\ref{R26}) can be further transformed. Namely, for the above values of $\alpha$ the integral in the second term in (\ref{R26}) converges absolutely and can be calculated by parts:
\begin{eqnarray}\label{R27}
& \displaystyle \frac{i}{\pi s} \! \int \!\d^2 {\bf q}' \;
\left[\frac{- s \alpha}{q'^{\,2}}\right]
\left(\frac{q^{\,2}}{q'^{\,2}}\right)^{\!\mbox{\scriptsize{i}} \alpha}  {\cal F}^2(q'^{\,2}) 
& \nonumber\\[0.5\baselineskip]
& \displaystyle = \; \left.
\left(\frac{q^{\,2}}{q'^{\,2}}\right)^{\!\mbox{\scriptsize{i}} \alpha}  {\cal F}^2(q'^{\,2}) \right|^{\infty}_{q'^{\,2}=0}
- \int^{\infty}_0 \!\d q'^{\,2} \;
\left(\frac{q^{\,2}}{q'^{\,2}}\right)^{\!\mbox{\scriptsize{i}} \alpha}  \left[ {\cal F}^2(q'^{\,2}) \right]' .&
\end{eqnarray} 
Here the prime after square brackets means differentiation with respect to $q'^{\,2}$. The first term in the r.h.s. (\ref{R27}) is zero for the specified $\alpha$ values. Substituting (\ref{R27}) into (\ref{R26}), we get a formula that allows us to return to the real $\alpha$,
\begin{eqnarray}\label{R28}
& \displaystyle 
F^{N+C}(s,q^2) \; e^{-\mbox{\scriptsize{i}} \alpha\ln (\lambda^2/q^2)} = \;-\; \frac{s \alpha}{q^2} \; {\cal F}^2(q^2) - \! 
\int^{\infty}_0 \!\!\!\d q'^{\,2} 
\left(\frac{q^{\,2}}{q'^{\,2}}\right)^{\!\mbox{\scriptsize{i}} \alpha} \! \left[ {\cal F}^2(q'^{\,2}) \right]'  F^{N}(s,q^2)
& \nonumber\\[0.2\baselineskip]
& \displaystyle - \; \frac{i\alpha}{\pi} \! \int 
\frac{\d^2 {\bf q}'}{q'^{\,2}} \,
\left(\frac{q^{\,2}}{q'^{\,2}}\right)^{\!\mbox{\scriptsize{i}} \alpha}   {\cal F}^2(q'^{\,2})
\left[F^{N}(s,[{\bf q-q}']^2) - F^{N}(s,q^2) \right]. &
\end{eqnarray}

Formula (\ref{R28}) represents the exact solution for the amplitude in the eikonal model \cite{Cahn}. We emphasize that we have obtained it in the case of a rapidly decreasing form factors at large $q^2$. Multiplying both sides in (\ref{R28}) by $\left( s/q^2 \right)^{\mbox{\scriptsize{i}} \alpha}$, we obtain an equivalent form of this solution with all the singularities at $q^2 \to 0$ collected in the Coulomb term,
\begin{eqnarray}\label{R29}
& \displaystyle \!\!\!\!\!
F^{N+C}(s,q^2) \, e^{-\mbox{\scriptsize{i}} \alpha\ln (\lambda^2/s)} = - \frac{s \alpha}{q^2} {\cal F}^2(q^2) \,
e^{\mbox{\scriptsize{i}} \alpha\ln (s/q^2)} \!- \!\! 
\int^{\infty}_0 \!\!\!\d q'^{\,2} 
\left(\frac{s}{q'^{\,2}}\right)^{\!\mbox{\scriptsize{i}} \alpha} \!\! \left[ {\cal F}^2(q'^{\,2}) \right]' \! F^{N}(s,q^2)
& \nonumber\\[0.2\baselineskip]
& \displaystyle - \; \frac{i\alpha}{\pi} \! \int 
\frac{\d^2 {\bf q}'}{q'^{\,2}} \,
\left(\frac{s}{q'^{\,2}}\right)^{\!\mbox{\scriptsize{i}} \alpha}   {\cal F}^2(q'^{\,2})
\left[F^{N}(s,[{\bf q-q}']^2) - F^{N}(s,q^2) \right]. &
\end{eqnarray}  
Note that for $\alpha = 0$, the second term in (\ref{R28}) and (\ref{R29}) is $F^{N}(s,q^2)$.

Having exact solution, we can obtain its expansion in powers of $\alpha$. As before, the result is ambiguous. If we proceed from (\ref{R28}), we come to the formula obtained in \cite{Cahn}:
\begin{eqnarray}\label{R30}
& \displaystyle 
F^{N+C}(s,q^2) \; e^{-\mbox{\scriptsize{i}} \alpha\ln (\lambda^2/q^2)} = \;-\; \frac{s \alpha}{q^2} \; {\cal F}^2(q^2) 
& \nonumber\\[0.2\baselineskip]
& \displaystyle + \; F^{N}(s,q^2) 
\left\{1 + \i \alpha \! \int^{\infty}_0 \!\d q'^{\,2} 
\ln\frac{q'^{\,2}}{q^{\,2}}  \left[ {\cal F}^2(q'^{\,2}) \right]' \right\} \!
& \nonumber\\[0.2\baselineskip]
& \displaystyle - \; \frac{i\alpha}{\pi} \! \int \d^2 {\bf q}' \;
\frac{{\cal F}^2(q'^{\,2})}{q'^{\,2}} \, 
\left[F^{N}(s,[{\bf q-q}']^2) - F^{N}(s,q^2) \right] +
O(\alpha^2) \,. &
\end{eqnarray} 
At the same time, based on (\ref{R29}), we arrive at the formula\footnote{It is curious to note that if we put ${\cal F}^2\!=\!1$ in (\ref{R30}) and (\ref{R31}), then formula (\ref{R21}) is restored. However, this coincidence is accidental, since the derivation of the above formulas was carried out in the case of a decreasing form factors.}
\begin{eqnarray}\label{R31}
& \displaystyle
F^{N+C}(s,q^2) \; e^{-\mbox{\scriptsize{i}} \alpha\ln (\lambda^2/s)} =  
- \, \frac{s \alpha}{q^2}  \; {\cal F}^2(q^2) 
& \nonumber\\[0.2\baselineskip]
& \displaystyle + \; F^{N}(s,q^2) 
\left\{1 + \i \alpha \! \int^{\infty}_0 \!\d q'^{\,2} 
\ln\frac{q'^{\,2}}{s}  \left[ {\cal F}^2(q'^{\,2}) \right]' \right\} \!
& \nonumber\\[0.5\baselineskip]
& \displaystyle - \; \frac{i\alpha}{\pi} \! \int \d^2 q' \;
\frac{{\cal F}^2(q'^{\,2})}{q'^{\,2}} \, 
\left[F^{N}(s,[{\bf q-q}']^2) - F^{N}(s,q^2) \right] +
O(\alpha^2) \,. &
\end{eqnarray}
The difference between (\ref{R30}) and (\ref{R31}) is $\,\i\alpha\ln(q^2/s) F^{N}(s,q^2)$, and this quantity is not vanishingly small in the general case. In particular, at $q^2=10^{-2}$ GeV$^2$ and $\sqrt{s} =8$--13 TeV, its absolute value is 25\% of the modulus of the nuclear amplitude, which translates into 6\% in the differential cross section. So, the $O(\alpha)$ amplitude approximation is unsuitable for processing data obtained with higher accuracy.

However, let us return to the exact solution. Actually, it can be further simplified by performing angular integration of either the nuclear amplitude or the form factor, as was proposed in \cite{Cahn} and \cite{KL}, respectively. Based on (\ref{R29}) in the former case we obtain
\begin{eqnarray}\label{R32}
& \displaystyle
F^{N+C}(s,q^2) \; e^{-\mbox{\scriptsize{i}} \alpha\ln (\lambda^2/s)} =  
- \, \frac{s \alpha}{q^2}  \; {\cal F}^2(q^2) \; 
e^{\mbox{\scriptsize{i}} \alpha\ln (s/q^2)} 
& \nonumber\\[0.2\baselineskip]
& \displaystyle 
- \int^{\infty}_0 \!\d q'^{\,2} 
\left( \frac{s}{q'^{\,2}} \right)^{\!\mbox{\scriptsize{i}} \alpha}
\left[ \widetilde{F}^{N}(s,q'^{\,2},q^2) {\cal F}^2(q'^{\,2}) \right]'  , &
\end{eqnarray} 
where
\begin{equation}\label{R33}
\widetilde{F}^{N}(s,q'^{\,2},q^2) = \frac{1}{2 \pi} \int^{2\pi}_0 \!\!\d \phi \, F^{N}(s,q^2 + 2 q q' \cos \phi + q'^{\,2})\,.
\end{equation}  
Note that due to $\widetilde{F}(s,0,q^2) = F(s,q^2)$ the second term in (\ref{R32}) becomes $F^{N}(s,q^2)$ at $\alpha = 0$. In the latter case, again based on (\ref{R29}), we get
\begin{eqnarray}\label{R34}
& \displaystyle
F^{N+C}(s,q^2) \; e^{-\mbox{\scriptsize{i}} \alpha\ln (\lambda^2/s)} =  
- \, \frac{s \alpha}{q^2}  \; {\cal F}^2(q^2) \; 
e^{\mbox{\scriptsize{i}} \alpha\ln (s/q^2)} - \, F^{N}(s,q^2) 
& \nonumber\\[0.2\baselineskip]
& \displaystyle \times
\left\{ \int^{\infty}_0 \!\!\!\d q'^{\,2} 
\left( \frac{s}{q'^{\,2}} \right)^{\!\mbox{\scriptsize{i}} \alpha} \!\! \left[ {\cal F}^2(q'^{\,2}) \right]'  
+ \i\alpha \! \int^{\infty}_0 \!\!\d q'^{\,2} \,
I(q'^{\,2},q^2)  
\left[\frac{F^{N}(s,q'^{\,2})}{F^{N}(s,q^2)} - 1 \right]\right\} , &
\end{eqnarray} 
where 
\begin{equation}\label{R35}
I(q'^{\,2},q^2) = \frac{1}{2 \pi} \int^{2\pi}_0 \d \phi \; 
\left(\frac{s}{q''^{\,2}}\right)^{\mbox{\scriptsize{i}} \alpha}
\frac{{\cal F}^2 (q''^{\,2})}{q''^{\,2}} \,,
\end{equation} 
and $q''^{\,2} \!= q^2 + 2 q q' \cos \phi + q'^{\,2}$. The second term in (\ref{R34}) at $\alpha \!=\! 0$ is $F^{N}(s,q^2)$, as well.

\section{Discussion and conclusion}\label{sec5}

The major result of this work is the exact solutions in the eikonal model \cite{Cahn} that describes CNI effect in elastic scattering of charged hadrons. We have shown that Coulomb-nuclear contributions to the amplitude are not exponentiated and cannot be added to the Coulomb phase. At the same time, due to the uncertainty of the overall complex phase in the exact solution after the elimination of IR divergences and the inhomogeneity of the amplitude in $\alpha$, its $O(\alpha)$-approximation is ambiguous. Moreover, this ambiguity is significant and can reach 25\% of the nuclear amplitude at $q^2=10^{-2}$ GeV$^2$ and $\sqrt{s} =8$--13 TeV, which translates into 6\% uncertainty in the differential cross section. This significantly exceeds the statistical errors of the data in the specified kinematic region \cite{TOTEM8,TOTEM13}, and means a significant decrease in the accuracy of determining the nuclear amplitude at low momentum transfers. At the same time, when using the exact solution the problem does not arise. The exact solution is represented in different forms in (\ref{R28}), (\ref{R29}) and (\ref{R32})--(\ref{R35}).

Nevertheless, there remains a problem with the applicability of the version of the eikonal model discussed above. Namely, in the case of spatially extended particles the question arises whether different parts of them can make individual Coulomb contributions. If the particles are scattered as integral objects, then the option with single Coulomb scattering must be realized. In this case the Coulomb contribution to the Born amplitude is represented by a single diagram with one-photon exchange, and the particular points of emission and absorption of the exchanged photon are probabilistic in nature and are determined by the probability distribution.\footnote{Recall that we use a QFT version of the eikonal model, in which interaction occurs due to the exchange of field quanta. This approach predetermines the method of introducing the form factors through averaging the amplitude operator by sandwiching it between the wave functions of scattering particles, see details in Sec.~\ref{sec3}.} On the contrary, if the different charged parts of the particles make separate contributions, then Coulomb scattering is formed by all scattering events. In this case the Born amplitude should be defined as a coherent sum of one-photon contributions. (This option will be discussed in details elsewhere.) Probably, this scattering mode can be effectively described in the potential eikonal approach, where the potential is a mean field that arises between systems of individual components. In this approach, electromagnetic form factors appear in exponential form directly in the eikonal phase. In fact, there are many works based on this approach, see e.g.~\cite{Islam1967,Petrov2018,Durand2020,Kaspar2021,Petrov2022} and section 4 in~\cite{Cahn}, but its validity in the case of scattering of protons is still unclear. 

In the general case, it is not known which mode is actually realized, with single or multiple Coulomb scattering. The result may depend on the internal structure of protons and on kinematic conditions of the scattering. Specifically in the CNI region, the single Coulomb scattering mode is more preferable. Really, in the case of very small transfers, such as $q^2 < 10^{-2} \,\mbox{GeV}^{2}$, where the Coulomb scattering is most imortant \cite{TOTEM8,Kaspar2011}, the wavelength of the exchanged photon is of the order of or greater than the transverse sizes of the colliding particles. Consequently, the time of the Coulomb interaction of the particles is of the order of or exceeds the time of interaction of their internal parts with each other. As a result, the internal parts have time to interact with each other during the time of emission or absorption of the photon. For this reason the particles participate in Coulomb scattering as a whole rather than as a collection of individual components, and strong correlation is not required to maintain this scattering mode. Of course, with increasing the momentum transfer, the wavelength of the exchanged photon decreases and the scattering should occur to a greater extent on individual parts of the particles. However, the Coulomb contributions become less significant in the higher transfer region and, therefore, the mentioned effect can remain with almost no observable consequences. 

In addition to the above discussion, we note that correlations between the internal parts of protons can increase with increasing energy of the collision. Such a behavior is predicted in the generalized parton model which takes into account the transverse parton motion \cite{Nekrasov2020,Nekrasov2021,Nekrasov2022}. In this model, the increase in the correlations occurs due to the strengthening of the coupling between the partons simultaneously with the accelerated growth with the energy of the transverse sizes of the protons. This effect manifests itself starting from energies 2--7 TeV, which is confirmed by the description of data \cite{Nekrasov2023}. 

To summarize, we note that in the general case the CNI problem in elastic proton scattering has not yet been completely solved. Within the framework of the eikonal approach, which is usually used for this purpose, we have found a complete solution (to all orders in $\alpha$) in an auxiliary case of scattering of point-like particles and in the case of spatially extended particles with single Coulomb scattering. The latter mode apparently corresponds to the real case of elastic scattering of protons at low momentum transfers at LHC energies.  

\bigskip

\noindent {\it Acknowledgments}: The author is grateful to V.A.~Petrov for stimulating discussions over the past few years.


\begin{thebibliography}{99}
\bibitem{Bethe}
H.~Bethe, Ann. Phys. 3 (1958) 190.
\bibitem{Soloviev}
L.D.~Solov'ev, JETP 22 (1966) 205.
\bibitem{WY}
G.B.~West, D.R.~Yennie, Phys. Rev. 172 (1968) 1413.
\bibitem{Cahn}
R.~Cahn, Z. Phys. C15 (1982) 253.
\bibitem{KL}
V.~Kundr\'{a}t and M.~Lokaji\v{c}ek, Z. Phys. C 63 (1994) 619.
\bibitem{Cheng-Wu}
H.~Cheng, T.T.~Wu,  Phys. Rev. 186 (1969) 1611.
\bibitem{Levy-Sucher}
M.~L\'{e}vy, J.~Sucher,  Phys. Rev. 186 (1969) 1656.
\bibitem{Collins}
P.D.B.~Collins, An introduction to Regge theory \& high energy physics, Cambridge Univ. Press, London. N.Y., Melbourne, 1977.
\bibitem{TOTEM8}
G. Antchev et al. (TOTEM Collaboration), Eur. Phys. J. C76 (2016) 661.
\bibitem{Gelfand}
I.M.~Gel'fand, G.E.~Shilov, Generalized Functions, vol. 1, Academic Press, New York, London, 1977.
\bibitem{TOTEM13}
G. Antchev et al. (TOTEM Collaboration), Eur. Phys. J. C79 (2019) 785.
\bibitem{Islam1967}
M.M.~Islam, Phys. Rev. 162 (1967) 1426.
\bibitem{Petrov2018}
V.A.~Petrov, Eur. Phys. J. C 78 (2018) 221 [Erratum, Eur.Phys.J. C 78 (2018) 414].
\bibitem{Durand2020}
L.~Durand, P.~Ha, Phys. Rev. D 102 (2020) 036025.
\bibitem{Kaspar2021}
J. Ka\v{s}par, Acta Phys. Pol. B 52 (2021) 85.
\bibitem{Petrov2022}
V.A.~Petrov, N.P.~Tkachenko, Phys.Part.Nucl. 54 (2023) 1152.
\bibitem{Kaspar2011}
J. Ka\v{s}par, V.~Kundr\'{a}t, M.~Lokaji\v{c}ek, and J.~Proch\'{a}zka, Nucl. Phys. B 843 (2011) 84.
\bibitem{Nekrasov2020}
M.L.~Nekrasov, Mod. Phys. Lett. A 35 (2020) 2050314.
\bibitem{Nekrasov2021}
M.L.~Nekrasov, Particles 4 (2021) 381. 
\bibitem{Nekrasov2022}
M.L.~Nekrasov,  Phys. Rev. D 106 (2022) 014028.
\bibitem{Nekrasov2023}
M.L.~Nekrasov, Phys. Rev. D 108 (2023) 034028.
\end{thebibliography}
\end{document}